\documentclass[]{spie}  
\usepackage[dvips]{graphicx}

\title{Real-Time Detection of Optical Transients with RAPTOR} 

\author{Konstantin Borozdin, Steven Brumby, Mark Galassi,
Katherine McGowan, Dan Starr,\\
W. Thomas Vestrand, Robert White, Przemyslaw Wozniak, and James Wren
\skiplinehalf
Los Alamos National Laboratory, Los Alamos, NM 87545 \\
}

\authorinfo{Send correspondence to K.Borozdin. E-mail: kbor@lanl.gov, Telephone:(1-505)665-1906}

\begin{document} 
  \maketitle 

\begin{abstract}
Fast variability of optical objects is an interesting though
poorly explored subject in modern astronomy.
Real-time data processing and identification of transient celestial
events in the images is very important for such study as it allows rapid
follow-up with more sensitive instruments. We discuss an approach which we
have developed for the RAPTOR project, a pioneering closed-loop system 
combining real-time transient detection with rapid follow-up.
RAPTOR's data processing pipeline is able to identify and localize 
an optical transient within seconds after the observation.
The testing we performed so far have been confirming the effectiveness
of our method for the optical transient detection. 
The software pipeline we have developed for RAPTOR
can easily be applied to the data from other experiments.
\end{abstract}



\section{INTRODUCTION}
\label{sect:intro}  

Over the last decade a substantial effort has been devoted to 
the development of astronomical optical telescopes capable
of a rapid response to gamma-ray bursts detected by space borne
instruments. A number of upper limits have been obtained in addition 
to a single truly spectacular discovery, a 9-magnitude optical flash
\cite{Akerlof99} associated with the GRB 990123, which was found to be 
at redshift z=1.6. This discovery has reaffirmed 
the importance of all-sky optical monitoring. There is
an acute scientific need
for an all-sky search\cite{Paczynski00} and early detection
of unexpected events, including GRB optical counterparts and
afterglows, supernovae, novae, dwarf novae, comets, asteroids,
gravitational microlensing events and more. 
Furthermore, about one million variable stars could be discovered 
and studied in detail by routine all-sky observations
with small wide-field optical telescopes. There are several ongoing projects 
which collect and archive the images for all the sky visible 
from a particular site every clear night\cite{Akerlof00,Park98,Pojmanski97}. 
However, none of these projects is able to detect
an optical transient in real time because of the lack of adequate
software. The wide field optical monitoring system RAPTOR (RAPid Telescope
for Optical Response\cite{Vestrand02}) is designed to identify 
and make follow-up observations of optical transients in real time.  
The most challenging aspect of the task has been the development
of a robust software able to do the job. In the following sections we 
briefly describe the project and discuss in more detail 
our software approach for the detection of optical 
transients in real time.

\section{RAPTOR: FIRST CLOSED-LOOP SYSTEM FOR OPTICAL ASTRONOMY}

\begin{figure}
   \begin{center}
   \begin{tabular}{c}
   \includegraphics[height=9cm]{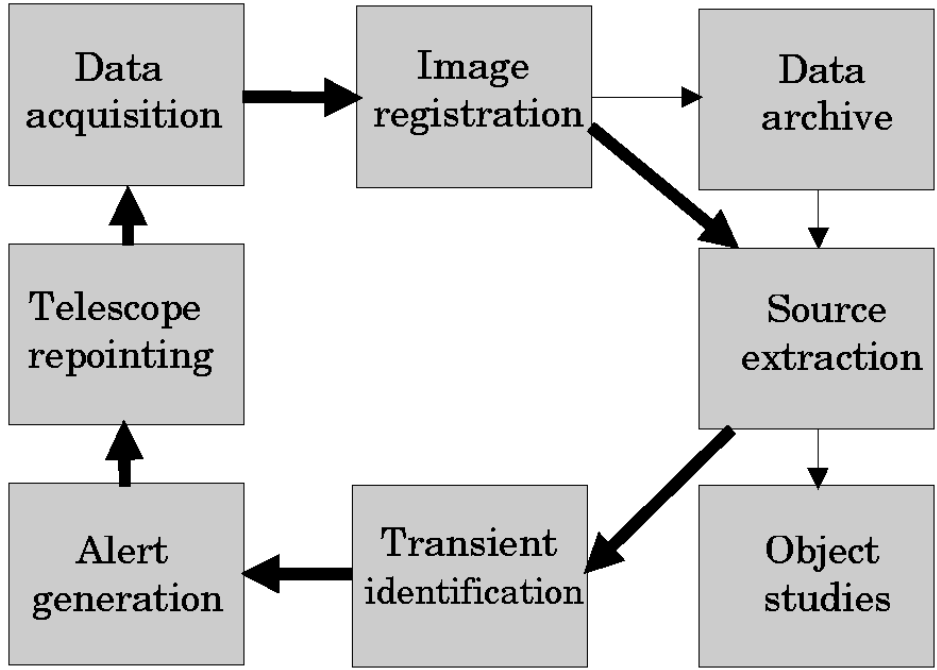}
   \end{tabular}
   \end{center}
   \caption[scheme] 
   { \label{fig:scheme} 
Flowchart of RAPTOR project. Thick solid lines represent tasks which should be 
completed in real-time (1 minute after data acquisition).
Thin lines denote later scientific analysis.
RAPTOR is the first closed-loop system for astronomy.
}
\end{figure}

The goal of the RAPTOR project is to detect an optical transient
within a wide-field of view, identify it automatically by
the real-time software pipeline, and perform follow-up
observations within the rapidly slewing narrow-field 
telescopes.  ``Zoological'' abbreviation of the project emphasizes
the concept of a system which is built as the analogy 
to biological vision systems.  Transient events are detected
with a wide-field telescopes which imitate peripheral vision.
For the follow-up  observations the transient is moved within field-of-view
of narrow-filed telescopes imitating fovea. Two fovea cameras
mounted on the common platform with wide-field instruments
and separate spectroscopic telescope will allow to track
the decline of an optical transient and to obtain its spectrum.
To our knowledge this is the first
experiment in optical astronomy which will be able both
to detect optical transients automatically and in real-time,
generate the alert and react to this alert for
follow-up observations.  To reach this ambitious goal we
need to build the closed loop system schematically
represented in Fig.\ref{fig:scheme}. Data acquisition, image
registration, source extraction, transient identification, alert
generation and telescope repointing are performed in real-time,
before an optical transient is likely to fade out.  
For optical counterparts of gamma-ray bursts this means
time scales on the order of one minute.

The real-time hardware/software pipeline is currently under testing for
the RAPTOR project.  The main components of the hardware are two
identical systems, Raptor-A and Raptor-B, combining wide-field
optical cameras (19 x 19 square degrees field-of-view) 
 with a more sensitive follow-up telescope (2 x 2 square degrees,
limiting magnitude m=17)\cite{Wren02}.
Data acquisition has been optimized for real-time
transient detection as discussed below.  Image registration
is performed as for other wide-field telescopes and is described
in detail elsewhere. Source extraction is based on freeware
packet SExtractor\cite{Bertin96}. SExtractor (Source-Extractor) is a program 
that builds a catalogue of objects from an astronomical image. 
It is particularly oriented towards reduction of large
scale galaxy-survey data, but it also performs well on moderately 
crowded stellar fields.  Both image registration and source extraction
were significantly sped up for the real-time RAPTOR pipeline.
Repointing of the telescope can be done extremely quickly 
due to the unique mounting of the RAPTOR telescopes capable
of slewing at a speed of about 100 degrees/s and accelerating to this 
velocity in less than one second\cite{Wren02}.
Transient identification and alert generation blocks are critical
for the operation of the whole pipeline and are newly developed
for the RAPTOR project.  So far the automatic detection of an optical 
transient has not been done by any astronomical project we are aware of.
In the following sections we shall concentrate on these parts of
the software pipeline.

\section{GENERAL APPROACH TO REAL-TIME DETECTION OF AN OPTICAL TRANSIENT 
WITH RAPTOR} 

The task of real-time transient detection was successfully solved
for X-ray all-sky monitoring about a decade ago\cite{Barthelmy94}.
However, detecting an optical transient is a more challenging task.
There are on the order of one hundred X-ray sources detectable in the whole 
sky by a modern X-ray monitor.  In comparison, the number of objects
obtainable by the RAPTOR wide-field telescopes is measured in 
the tens of millions.
Many of these are variable sources, that are difficult to distinguish
from optical transients.  There are a lot 
of sources in the images which can imitate optical transients: 
flaring stars, comets, asteroids, meteorites, satellites, airplanes, 
hot pixels and image defects.  Many false positives 
can not be distinguished from the event of interest based on
a single image only.  This makes it essential to use additional
information, such as previous observations, parallax measurements
with the second telescope, matching objects in consecutive observations
etc.

Our knowledge of the optical counterparts of gamma-ray bursts is limited. 
So far, only one event has been observed\cite{Akerlof99}.  
What is expected is an optical flash from an ``empty'' region 
of the sky which fades out on time scale of few minutes.  
To know which areas of the sky are ``empty'' we need to compare 
current observation with previous observations of the same field, 
either in the form of an image or a source catalog.  
To be able to detect a variation we need to have at least two,
or better three, consecutive images of the same field. 
The expected time scales for variability of GRB optical counterparts
are order of one minute, which dictates that a single exposure
should be shorter than this time.  But as we discussed above
a single exposure is not enough.  So we have designed an observational
sequence, which consists of multiple consecutive 30-sec
exposures of the same field in the sky. As the distribution of GRBs
is isotropic\cite{BATSE92} the probability of detection for 
a GRB optical counterpart depends on the size of telescope field-of-view 
and is independent from the pointing direction.  However, we should try
to avoid crowded fields in the Galactic plane where the effective 
area of ``empty'' regions is limited by the abundance of 
known sources. Visibility constraints dictate the choice of
the field close to the zenith direction.

   \begin{figure}
   \begin{center}
   \begin{tabular}{c}
   \includegraphics[height=9cm]{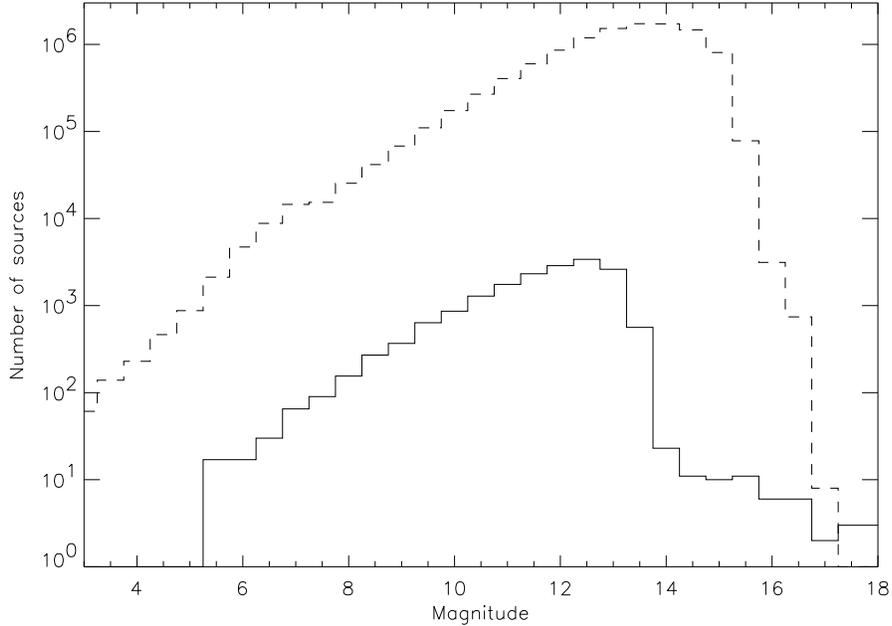}
   \end{tabular}
   \end{center}
   \caption[catmag] 
   { \label{fig:catmag} 
Magnitude distribution of sources detected in a single image with 
RAPTOR wide-field camera (camera C on Raptor-A system, 60 s
exposure of the field of Regulus). 
For comparison the magnitude distribution of sources in  
Guide Star Catalog is shown (dashed line).
Completeness limit is around m=13 for GSC.
}
\end{figure}

Another important feature of the RAPTOR project is stereoscopic
observations of the same field in the sky. This is achieved by
having two separate optical systems (Raptor-A and Raptor-B) 
with a separation of 38 km between them.  
This approach allows us to exclude all local objects 
contributing to false positives from the analysis.
Comparison of two simultaneous observations with two different
instruments will allow one to reduce the number of hot pixels among
the false positives.  To reduce the number of matched hot pixels
in consecutive observations with the same telescope we
apply dithering of the pointing direction, i.e. the center
of the field of view is shifted by a fraction of a degree
for two consecutive observations of the same sky field.

\section{CATALOG}

   \begin{figure}
   \begin{center}
   \begin{tabular}{c}
   \includegraphics[height=9cm]{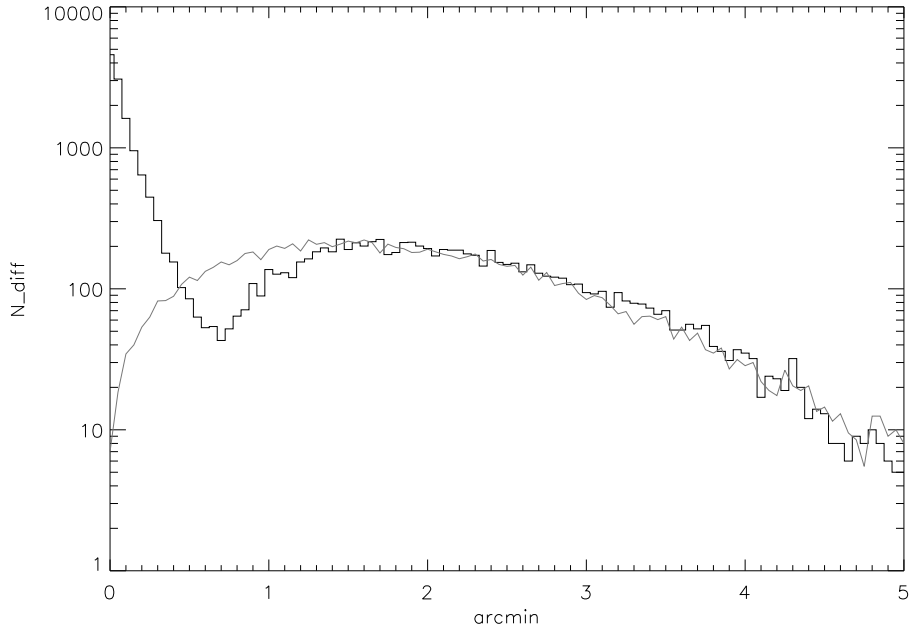}
   \end{tabular}
   \end{center}
   \caption[neighbors] 
   { \label{fig:neighbors} 
Distribution of number of detected sources versus the distance 
to the closest match. Thin grey line shows random distribution 
of sources, solid histogram represents experimental data 
for two ROTSE observations of the same field. 
Rapidly declining component on the left is composed of
identical sources, detected in both observations, 
broad component are random coincidences.
}
\end{figure} 

To understand which objects are ``new'' in the image,
the system should use the information from previous observations.
In the RAPTOR pipeline we compare the new source list with 
the catalog of known sources and weed out matches.
The approach is quite obvious, but its implementation
is a bit trickier.  First, you need a catalog which is complete 
for the sensitivity limit of the RAPTOR telescopes,
but does not contain too many fainter sources, which would 
reduce an affective area for a transient detection.
Also, available astronomical catalogs are static (i.e. lack
time dimension).  As a result they do not include 
many objects which demonstrate dramatic variability like
novae, pulsating and cataclysmic variables. We tested 
the use of the GSC (Guide Star Catalog\cite{Lasker90}, 
with ROTSE (Robotic Optical Transient Search Experiment) data 
and found that even though the limiting sensitivities 
are comparable, many ROTSE sources were unmatched in the GSC 
(see Table\ref{tab:criteria}).
More promising is the use of an updateable self-produced catalog
obtained with the same instrument based on the results of 
previous observations. In this case we start from the GSC and 
expand it with objects matched in consecutive observations. 
This approach has been successfully tested with ROTSE data. 
For RAPTOR wide-field cameras most of the sources are expected
to be present in GSC (see Fig.\ref{fig:catmag}), but for more
sensitive fovea cameras just small fraction of the detected sources
can be found there.  In the RAPTOR project we will keep and update 
both more shallow catalog for wide-field cameras and deeper catalog
for follow-up fovea observations.  We expect that the former 
catalog will contain on the order of a million objects while
the latter one may eventually include as many as 50 million
objects.

An important parameter is match radius, i.e. the maximum difference
in coordinates of two objects detected in two different observations 
which are still considered to be the same source in the sky. To define 
this parameter for ROTSE test data we have obtained the distribution 
of the distance to the nearest neighbor (Fig.\ref{fig:neighbors}).  
Two components, one composed of identical sources detected 
in two separate observations and the other which consists
of random coincidences are easily distinguished. The exact choice 
of match radius depends on how many false positives you allow
to be kept and what percentage of real matches you allow to be rejected.

Updateable source catalogs from the RAPTOR observations will play 
important role in the real-time detection of optical transients.
Also they will allow more detailed variability studies 
for large group of interesting objects and will supply
the information for SkyDOT sky database\cite{Wozniak02}.

\section{IMAGE QUALITY CHECKS}

\begin{figure}
   \begin{center}
   \begin{tabular}{c}
   \includegraphics[height=9cm]{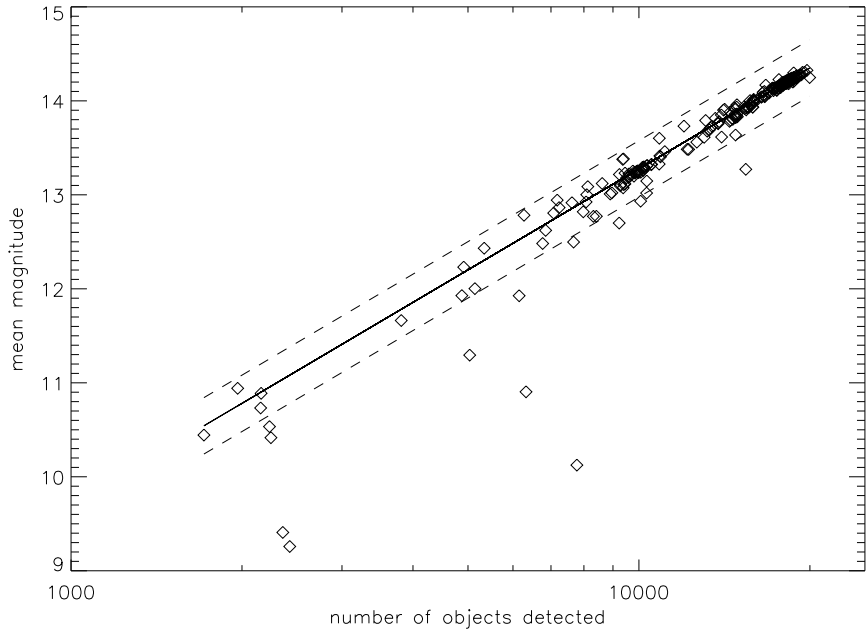}
   \end{tabular}
   \end{center}
   \caption[imaq] 
   { \label{fig:imaq} 
Dependence of mean magnitude of the detected sources
on their number.  Each diamond represent one test image of 
the same field. We found empirically that most of the
good quality images lie along the straight line in logarithmic Al
scale. We accepted the images which lie within the two dashed
lines and rejected outliers.
}
\end{figure}

An essential component of a fully automatic system for 
optical transient detection is an image quality check
which allows one to exclude from the analysis the images
which have substantial defects due to weather conditions
during the observation, or malfunction of some
hardware components.  We developed several simple checks
based on empirically found differences between good and bad
images which could be easily included into the real-time
analysis. Fig.\ref{fig:imaq} illustrates one of the checks.
For each test image we calculated the number of the detected
sources and their mean magnitude for a given field.
We accepted images which lie close to the straight line in 
this plot and rejected outliers, which appeared to have 
significant image defects. In our analysis we also filtered 
the images according to the total number of detected sources
and percentage of the sources matched with the catalog.

It is also important in automatic image analysis,
especially when you would like to detect source variability,
to reconstruct reliably the brightness of each source.
However, the measured magnitude of the sources is affected
not only by the quality of the whole image but also by local
image defects, which are particularly important for wide-field
telescopes, where separate parts of the field of view
may have very different seeing conditions.  To correct for local
defects we use a relative photometry method.
The idea is to compare the mean measured magnitude
of a large group of closely located sources, with
the same group of sources from a standard image or source
catalog.  Significant deviation of the mean measured magnitude from 
the nominal value indicates that there is some problem with this part 
of the image and the magnitude of each source in the region 
needs to be corrected.  In the case of a large value
deviation for the whole image or a large part of the image, then
this image or part of the image should be excluded from analysis.

   \begin{figure}
   \begin{center}
   \begin{tabular}{c}
   \includegraphics[height=9cm]{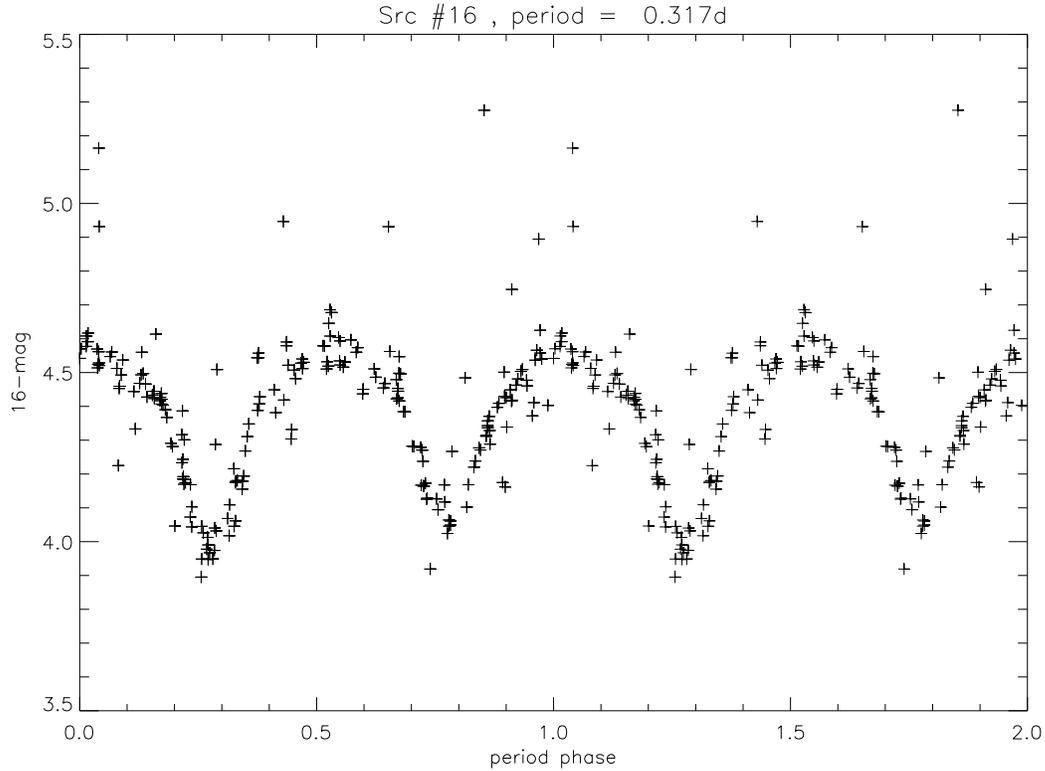}
   \end{tabular}
   \end{center}
   \caption[rel_phot] 
   { \label{fig:lc_raw} 
Folded light curve of a variable star obtained with
automatic RAPTOR pipeline without photometry correction
(crosses).
}
\end{figure} 

   \begin{figure}
   \begin{center}
   \begin{tabular}{c}
   \includegraphics[height=9cm]{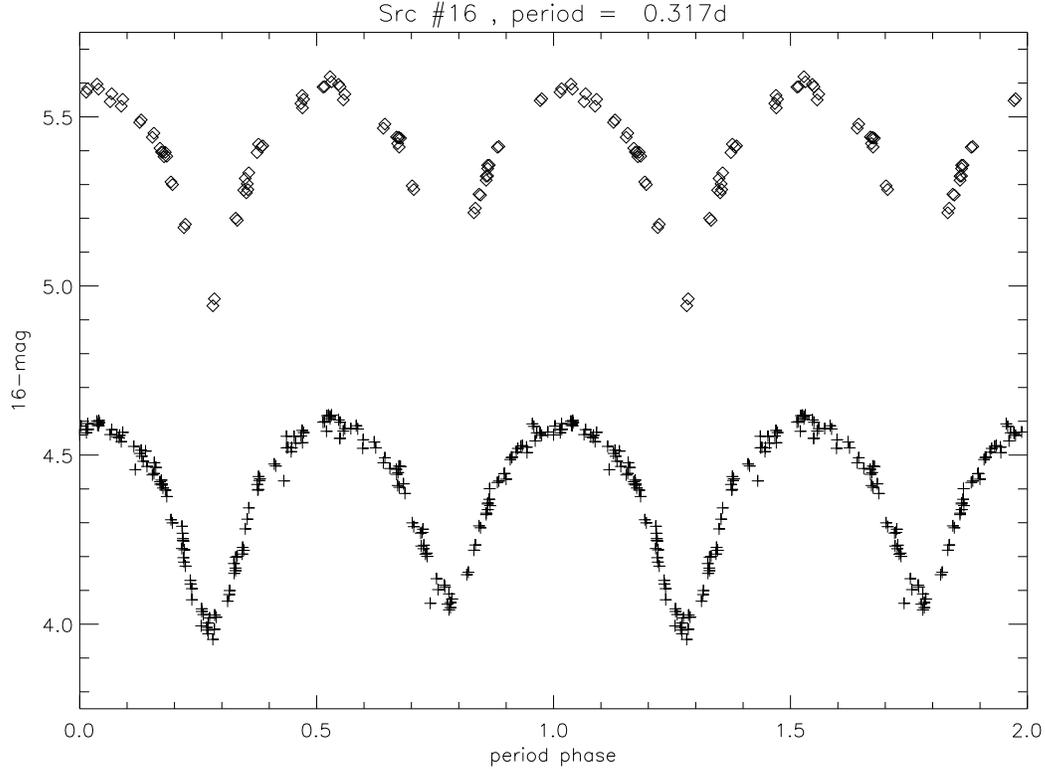}
   \end{tabular}
   \end{center}
   \caption[rel_phot] 
   { \label{fig:rel_phot} 
Folded light curves for the same variable star 
as shown in Fig.\ref{fig:lc_raw}. Diamonds represent data
obtained by human screening of images\cite{Akerlof00}.
Crosses are produced by the RAPTOR pipeline with application
of automatic image quality control criteria and
relative photometry correction.  The diamonds are shifted along 
vertical axis for clarity. The crosses are obtained from
an analysis of larger set of observations.
}
\end{figure}

The result of the application of automatic image quality checks
and relative photometry correction to ROTSE
observations is illustrated in Figs.\ref{fig:lc_raw} and \ref{fig:rel_phot}.
Fig.\ref{fig:lc_raw} represent the light curve for a variable
star\cite{Akerlof00} produced by automatic RAPTOR pipeline.
Fig.\ref{fig:rel_phot} shows the same light curve
but obtained with automatic image quality checks
and relative photometry correction included into the pipeline.
For comparison we also show the light curve for the same 
variable star\cite{Akerlof00} produced by a human expert.
The comparison shows that quality of the data produced
by the RAPTOR pipeline is comparable with the results of 
data analysis performed by the scientist.

\section{ALERT GENERATION}

\begin{figure}
   \begin{center}
   \begin{tabular}{c}
   \includegraphics[height=11cm]{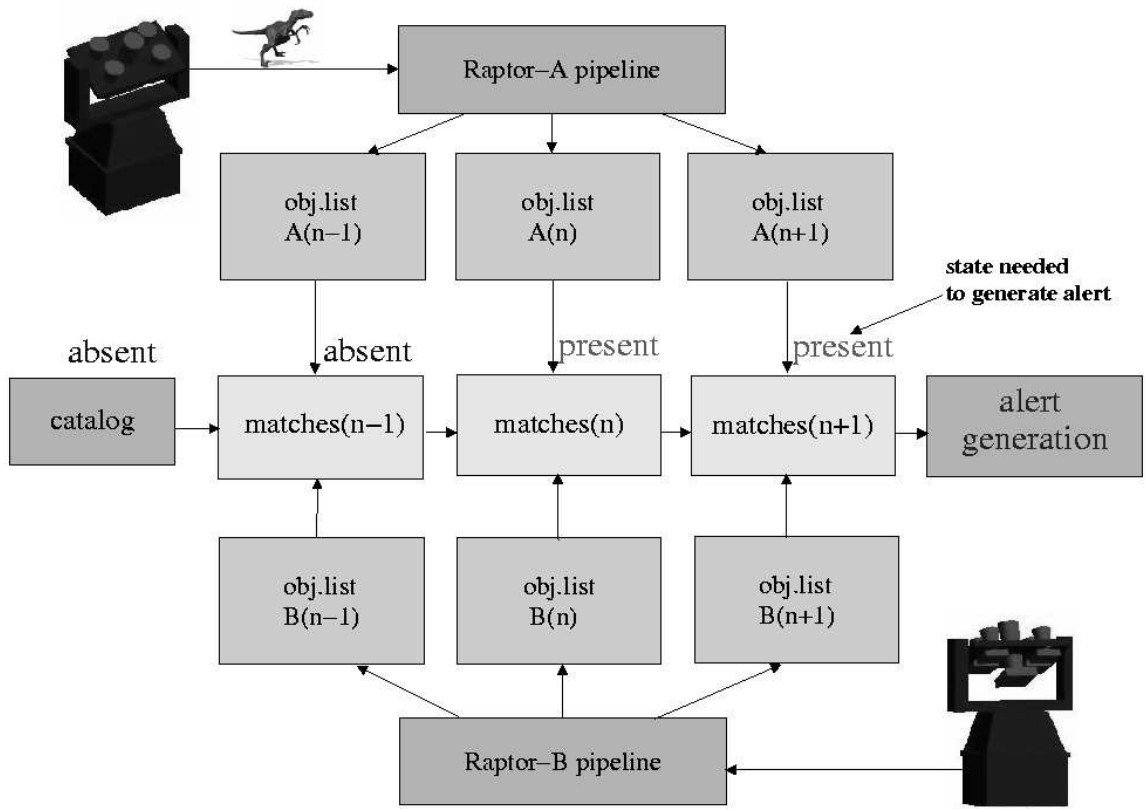}
   \end{tabular}
   \end{center}
   \caption[alerts] 
   { \label{fig:alerts} 
Alert generation for an optical transient found with the
RAPTOR pipeline.  Two identical systems, Raptor-A and
Raptor-B, observe the same sky field from two different locations
on the ground. Images are analyzed and object lists are
generated each 30 s.  The sources in the object lists are matched
for the two lists obtained from different telescopes. 
Then the lists for two consecutive exposures are matched.
The matched sources in two consecutive exposures
(n and n+1) are compared with previous exposure (n-1) and with
the catalog of previous observations.  If the source
present in exposures n and (n+1), but was not detected neither 
in previous exposure, nor in the catalog of previous observations,
an alert is generated.
}
\end{figure}

Fig.\ref{fig:alerts} shows the scheme of alert generation
by the RAPTOR system.  The alert generation is optimized 
for the search of optical counterparts of gamma-ray bursts.
The same area of sky is observed with two identical wide-field
telescopes installed in two separate positions (with a separation
of 38 km between them). The duration of each exposure
is 30 s.  Pointing direction slightly changes 
(dithers) between the exposures. Each image is analyzed 
immediately after its acquisition and a list of detected objects 
is built for each exposure and each camera.  The list obtained 
with Raptor-A is compared with the list of objects from Raptor-B image, 
and a list of matches is produced.  At this stage most of the hot pixels,
image defects and local objects, which will have significant
parallax in the images obtained from the two telescopes, are rejected.
Next we compare results from two consecutive exposures
which allows one to eliminate random coincidences in the images.
A new list of matches is compared with the previous exposure, (n-1),
which has been analyzed at the previous step of the procedure.
If the object was present in the (n-1) list of matches then 
it is not a new object for exposure n and should be rejected
from the list of potential alerts.  Comparison to the catalog
allows to eliminate constant and slowly variable sources
which for whatever reason may not be detected in the exposure (n-1).
If the list of potential alerts is not empty after 
the completion of this step then an alert is generated and
the source will be observed with more sensitive follow-up
telescopes to follow its evolution down to about 17th magnitude 
and improve the localization of the object from fraction 
of arc minute for initial detection with a wide-field camera 
to few arcsec for follow-up observations with fovea telescopes.

\section{TEST RESULTS}

The algorithm of optical transient detection 
has been coded in the C computer language and tested under 
Linux operational system with the sequence of ROTSE-I data.
ROTSE-I telescopes are similar to the wide-field telescopes
of the Raptor-A and Raptor-B systems.  However, the observational
sequence we have used in the testing 
has not been optimized for the search of optical
transients according to our method.  Using these data we could
not expect to detect any fast optical transients, only
long-living transients which are bright for time scales longer
than a day. Of course, we could not
utilize simultaneous observations of the same field with
two different telescopes which is an unique capability of
RAPTOR system.  Still the results were quite promising.
We have been able to suppress the rate of false positives
down to 1 per 50 observations (see Table\ref{tab:criteria}
for more details).  

\begin{table}[h]
\caption{Reduction of number of false positives in the test data
by application of different filtering criteria (representative numbers
for a sequence of testing images)}
\label{tab:criteria}
\begin{center}       
\begin{tabular}{|l|l|} 
\hline
\rule[-1ex]{0pt}{3.5ex}  Filtering criteria & Average number of objects / image  \\
\hline
\hline
\rule[-1ex]{0pt}{3.5ex}  Initial image (no filtering) & 15,500 objects / image\\
\hline
\rule[-1ex]{0pt}{3.5ex}  Matching with GSC & 2,600 unmatched sources / image\\
\hline
\rule[-1ex]{0pt}{3.5ex}  Matching with internal updateable catalog & 460 unmatched sources / image\\
\hline
\rule[-1ex]{0pt}{3.5ex}  Pair matching of sources in consecutive images & 85 false positives / image \\
\hline
\rule[-1ex]{0pt}{3.5ex}  Ditherings between observations & 0.7 false positives / image \\
\hline
\rule[-1ex]{0pt}{3.5ex}  Brightness level trigger (13th magnitude min) & 0.02 false positives / image\\
\hline
\rule[-1ex]{0pt}{3.5ex}  Stereoscopic observations & not tested yet \\
\hline 
\end{tabular}
\end{center}
\end{table}

This rate may be acceptable
for internal alerts, but we expect that using
stereo observations we shall be able to eliminate most
of the remaining false positives.  The typical false positive
during our tests was a very faint source which barely exceeds
the detection limit and so was not included into the catalog.
By random coincidence a hot pixel overlaps the position of this source
in one of observations.  As a result we detect a strongly variable
source in two consecutive exposures and generate the alert.
Another type of false positive is random coincidence of two different 
hot pixels.  The number of these and similar false positives will be 
further significantly reduced in stereo RAPTOR observations.
During the tests on standard commercially available PCs 
the pipeline has been able to process 
the data in a rate sufficient for real-time processing 
of the RAPTOR data. 

\section{CONCLUSION}

We have designed and developed a software pipeline
for a real-time detection of astrophysical transients in optical
observations.  The pipeline has developed for the RAPTOR project 
which is the first closed-loop system for optical astronomy
combining real-time transient detection with rapid follow-up
capabilities.  The testing of the pipeline with ROTSE data
has confirmed that our approach can be effective for the search
of optical transients.  Applying various filtering 
criteria we were able to reduce the number of false positives 
for transient alerts by several orders of magnitude.
We also demonstrate that automatic image
quality checks and relative photometry corrections can be applied
with the pipeline to obtain automatically the results suitable
for scientific data analysis. Quality of these results is comparable
with the performance of a human expert. The speed of 
the pipeline is adequate for the processing of the RAPTOR data
in real-time and generate alerts within 1 minute after the data
acquisition.

We plan to perform a full-scale testing of the pipeline
with the RAPTOR data in the coming months.

\acknowledgments     
 
Internal Laboratory Directed Research and Development
funding supports the RAPTOR project at Los Alamos 
National Laboratory under DoE contract W-7405-ENG-36.
Los Alamos National Laboratory is operated by the University 
of California for the National Nuclear Security
Administration (NNSA) of the U.S. Department of Energy and works 
in partnership with NNSA's Sandia and Lawrence
Livermore national laboratories to support NNSA in its mission.
This research has made use of data provided by ROTSE\cite{Akerlof99}
robotic telescopes.

\bibliography{OT}   

\begin{thebibliography}{10}

\bibitem{Akerlof99}
C.~Akerlof, R.~Balsano, S.~Barthelemy, J.~Bloch, P.~Butterworth, D.~Casperson,
  T.~Cline, S.~Fletcher, F.~Frontera, G.~Gisler, J.~Heise, J.~Hills, R.~Kehoe,
  B.~Lee, S.~Marshall, T.~McKay, R.~Miller, L.~Piro, W.~Priedhorsky,
  J.~Szymanski, and J.~Wren, ``Observation of contemporaneous optical radiation
  from a gamma-ray burst,'' {\em Nature} {\bf 398}, pp.~400--402, 1999.

\bibitem{Paczynski00}
B.~Paczynski, ``Monitoring all sky for variability,'' {\em The Publications of
  the Astronomical Society of the Pacific} {\bf 112}, pp.~1281--1283, 2000.

\bibitem{Akerlof00}
C.~Akerlof, S.~Amrose, R.~Balsano, J.~Bloch, D.~Casperson, S.~Fletcher,
  G.~Gisler, J.~Hills, R.~Kehoe, B.~Lee, S.~Marshall, T.~McKay, A.~Pawl,
  J.~Schaefer, J.~Szymanski, and J.~Wren, ``{ROTSE} all-sky surveys for
  variable stars. {I}. {T}est fields,'' {\em The Astronomical Journal} {\bf
  119}, pp.~1901--1913, 2000.

\bibitem{Park98}
H.-S. Park, E.~Ables, S.~Barthelmy, D.~Scott, R.~Bionta, L.~L. Ott, E.~Parker,
  and G.~Williams, ``Instrumentation of {LOTIS} -- {L}ivermore {O}ptical
  {T}ransient {I}maging {S}ystem: a fully automated wide-field-of-view
  telescope system searching for simultaneous optical counterparts of gamma-ray
  bursts,'' in {\em Optical Astronomical Instrumentation},  S.~D'Odorico, ed.,
  {\em Proc. SPIE} {\bf 3355}, pp.~658--664, 1998.

\bibitem{Pojmanski97}
G.~Pojmanski, ``The all sky automated survey,'' {\em Acta Astronomica} {\bf
  47}, pp.~467--481, 1997.

\bibitem{Vestrand02}
W.~Vestrand, K.~Borozdin, S.~Brumby, D.~Casperson, E.~Fenimore, M.~Galassi,
  G.~Gisler, K.~McGowan, S.~Perkins, W.~Priedhorsky, D.~Starr, , R.White,
  P.~Wozniak, and J.~Wren, ``Searching for optical transients in real-time. the
  {RAPTOR} experiment,'' in {\em Gamma-Ray Bursts and Afterglow Astronomy},
  D.~Kocevsky, F.~Ryde, M.~Boettcher, and I.~Smith, eds., {\em Proc. of the
  Woodshole GRB Conference}, 2002 (in press).

\bibitem{Bertin96}
E.~Bertin and S.~Arnouts, ``S{E}xtractor: Software for source extraction,''
  {\em Astronomy and Astrophysics Supplement} {\bf 117}, pp.~393--404, 1996.

\bibitem{Wren02}
J.~Wren, K.~Borozdin, S.~Brumby, M.~Galassi, K.~McGowan, D.~Starr, W.~Vestrand,
  R.White, and P.~Wozniak, ``A distributed control system for rapid
  astronomical transient detection,'' in {\em Advanced Global Communications
  Technologies for Astronomy},  {\em Proc. of SPIE} {\bf 4845-21}, 2002 (this
  conference proceedings).

\bibitem{Barthelmy94}
S.~D. Barthelmy, T.~Cline, N.~Gehrels, T.~Bialas, M.~Robbins, J.~Kuyper,
  G.~Fishman, C.~Kouveliotou, and C.~Meegan, ``{BACODINE}: The real-time
  {BATSE} gamma-ray burst coordinates distribution network,'' in {\em Gamma-Ray
  Bursts},  G.~J. Fishman, ed., {\em AIP Conference Proceedings} {\bf 307},
  p.~643, 1994.

\bibitem{BATSE92}
C.~Meegan, G.~Fishman, R.~Wilson, J.~Horack, M.~Brock, W.~Paciesas, G.~N.
  Pendleton, and C.~Kouveliotou, ``Spatial distribution of gamma-ray bursts
  observed by {BATSE},'' {\em Nature} {\bf 355}, pp.~143--145, 1992.

\bibitem{Lasker90}
B.~Lasker, C.~Sturch, B.J., J.~Russell, H.~Jenkner, and M.~Shara, ``The {G}uide
  {S}tar {C}atalog. {I} - {A}stronomical foundations and image processing,''
  {\em Astronomical Journal} {\bf 99}, pp.~2019--2058, 2173--2178, 1990.

\bibitem{Wozniak02}
P.~Wozniak, K.~Borozdin, M.~Galassi, W.~P.~D. Starr, T.~Vestrand, R.~White, and
  J.~Wren, ``Virtual observatory for variable stars study,'' in {\em Virtual
  Observatories},  {\em Proc. of SPIE} {\bf 4846-25}, 2002 (this conference
  proceedings).

\end{thebibliography}
\bibliographystyle{spiebib}   

\end{document}